# Discovery of a Magnetic Topological Semimetal Eu$_3$In$_2$As$_4$ with a Single Pair of Weyl Points

Ke Jia[1,2,3#], Jingyu Yao[1,3#], Xiaobo He[4#], Yupeng Li[1,2,3#], Junze Deng[1,3], Ming Yang[4], Junfeng Wang[4], Zengwei Zhu[4], Cuixiang Wang[1], Dayu Yan[1], Hai L. Feng[1,5], Jie Shen[1,2,3,5*], Yongkang Luo[4*], Zhijun Wang[1,3*], Youguo Shi[1,2,3,5*]

[1]*Beijing National Laboratory for Condensed Matter Physics, Institute of Physics, Chinese Academy of Sciences, Beijing 100190, China*

[2]*Center of Materials Science and Optoelectronics Engineering, University of Chinese Academy of Sciences, Beijing 100049, China*

[3]*School of Physical Sciences, University of Chinese Academy of Sciences, Beijing 100190, China*

[4]*Wuhan National High Magnetic Field Center and School of Physics, Huazhong University of Science and Technology, Wuhan 430074, China*

[5]*Songshan Lake Materials Laboratory, Dongguan, Guangdong 523808, China*

\# These authors contributed equally.

# Abstract.

Magnetic Weyl semimetal (MWS) is a unique topological state with open surface Fermi arc states and other exotic transport phenomena. However, most reported MWSs show multiple pairs of Weyl points and complicated Fermi surfaces, which increases the difficulty of the investigation into the intrinsic chiral transport property. In this work, we successfully synthesized a soft magnetic Weyl semimetal $Eu_3In_2As_4$ with a single pair of Weyl points under magnetic fields. The Shubnikov–de Haas (SdH) oscillation with a single frequency, as well as a linear hall resistance with the same carrier density, is observed up to 50 Tesla, indicating a single pair of Weyl points around the Fermi level with a massless fermion ($m^* = 0.121\ m_0$, $\pi$ Berry phase). Such a single pair of Weyl points is further confirmed by the density functional theory calculations. The magnetic ordering and band topology can be easily tuned by the external magnetic field. The field-induced MWS $Eu_3In_2As_4$ with a single pair of Weyl points is a good platform to detect chiral transport properties, including possible quantum anomalous Hall effect.

# Introduction

Weyl semimetals characterized by the Weyl point have quickly gained popularity since their discovery. Many Weyl semimetals are experimentally reported, such as TaAs[1–4], TaP[5,6], WTe$_2$[7], and SrMnSb$_2$[8]. Weyl points with opposite chirality connected by an open Fermi arc and chiral edge modes can result in the anomalous Hall effect [9–11]. Furthermore, MWSs attract much attention for their quantum anomalous Hall effect (QAHE) [8,12–19]. In particular, the QAHE characterized by dissipationless transport of chiral edge states is one of the fascinating states in MWSs due to the potential application for new-generation electronic devices[15–20]. Many efforts have been devoted to discovering new types of MWSs to realize QAHE[15–19]. However, most reported ones show complicated Fermi surfaces with multiple pairs of Weyl points[10,21,22], compromising the study of their chiral transport features. One candidate with single pair of Weyl points is the ferromagnetic MnBi$_2$Te$_4$ [23–27], which has been reported in Ref[24].

To explore MWSs, one possible way is to substitute magnetic ions into non-magnetic insulators. For example, when substituting Sr with Eu for insulators SrIn$_2$P$_2$ and SrCd$_2$As$_2$[28,29], the resulting EuIn$_2$P$_2$ and EuCd$_2$As$_2$ are MWSs [30,31]. Sr$_3$In$_2$As$_4$ has been predicted to be an insulator with an energy gap of up to 0.3 eV.[32] In this work, we anticipated substituting Sr with Eu into Sr$_3$In$_2$As$_4$, and high-quality single crystals of Eu$_3$In$_2$As$_4$ are successfully synthesized. Density-functional theory (DFT) results reveal a single pair of Weyl points around the Fermi level under the ferromagnetic (FM) state, indicating that Eu$_3$In$_2$As$_4$ is a MWS candidate. Electron transport results show a single-frequency Shubnikov–de Haas effect (sdH) oscillation and linear Hall resistance. Analysis of sdH oscillation and Hall resistance yield similar carrier densities, and nontrivial π Berry phase is derived by Landau Level (LL) fan diagram and Lifshitz–Kosevich(L-K) formula, supporting the single pair of Weyl points. Magnetization proves the intimate relationship between magnetic state and Weyl semi-metallic state. In particular, high mobility up to 3900 cm$^2\cdot$(v$\cdot$s)$^{-1}$ and sdH oscillation surviving at 75 K may indicate the potential realization of QAHE in thin film of the system at high temperature.

# Result and discussion

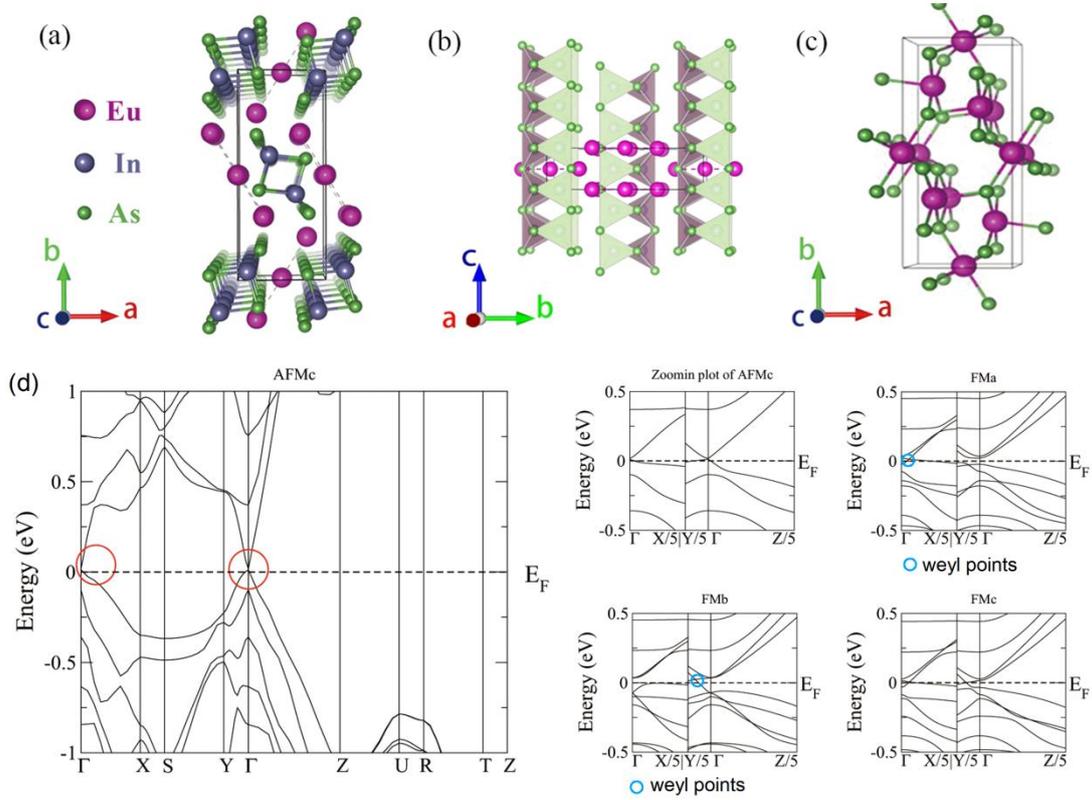

Fig.1 (a) and (b) represent the crystal structure viewed from the *c*- and *a*-axis, respectively. (c) Two types of Eu atoms coordinated with 4 and 6 As; (d) The band structures of $Eu_3In_2As_4$ for the four magnetic configurations (AFMc, FMc, FMa, FMb). The black dashed line represents the Fermi energy. The gap of the AFM configuration is 4.4 meV at Γ, while it becomes negative in the FM configurations. The distribution of Weyl/gapless points depends on the FM directions.

*Magnetic configurations and result of first-principles calculations.* Details of calculation and method of experiment on $Eu_3In_2As_4$ are shown in SI 4 and 10. $Eu_3In_2As_4$ crystallizes in orthorhombic structure with space group (SG) *Pnnm* (No. 58). One-dimensional chains [$In_2As_4$] aligns along the *c*-axis by edge-sharing. The corresponding magnetic space groups (MSG) of $Eu_3In_2As_4$ (#471-477 shown in SI.3) are considered in TopMat[33]. The magnetic configurations are presented explicitly in SI.4. Among these seven configurations, only three of them, 471 c-direction antiferromagnet (AFMc), 475 c-direction ferromagnet (FMc) and 476 (canted-FM,

FMa, FMb), are consistent with the state Eu$^{2+}$ (the magnetic moment of some Eu is confined to zero in other MSG). Then, we performed the non-collinear moment DFT calculations with Hubbard U = 7 eV for the Eu-f electrons. The converged total energies and band structures are obtained and shown in Table.1 and Fig.1(d), respectively. Please find their detailed configurations in SI 4. The SG number, MSG number, and magnetic types are given in the first three columns, which are not valid for the cases of NM and ZM. "NM" represents nonmagnetic (type-II) MSG, "ZM" represents some magnetic atoms with zero moments required by the given MSG. The calculations of magnetic configurations with symmetry indicators (SIs) are performed under the consideration of spin-orbital coupling. We find that the band topology varies for different magnetic configurations, although their total energies are almost the same (less than 0.1 meV; indicating a soft magnet). In AFMc, there is a simple band structure and the tiny band gap is trivial at Γ. The highest valence states (HVS) are from the As-*s* orbitals, and the lowest conduction states (LCS) are mainly from the Eu-*d* orbitals. In FMa/FMb/FMc, it becomes a semimetal with a negative band gap and the hybridization between HVS and LCS gives rise to a semimetal surface, as the compatibility relations (CR) conditions are not satisfied. Our DFT calculation show it is a magnetic nodal line semimetal in FMc configuration (the nodal line in the XY plane), while it is a magnetic Weyl semimetal in FMa/FMb configuration (the single pair of Weyl points on the ΓX/ΓY line), which are indicated in Fig.1 (d).

Table I. The list of results of SG 58 $Eu_3In_2As_4$ given by TopMat

(http://tm.iphy.ac.cn/TopMat_1651msg.html)[33].

|  | MSG(#OG) | Type | Configuration | Energy(eV/atom) | SIs |
|---|---|---|---|---|---|
| $Eu_3In_2As_4$ SG58 | 471 | I | AFMc | -6.9430 | $Z_2$=0 |
|  | 472 | II | NM | ∅ | ∅ |
|  | 473 | III | ZM | ∅ | ∅ |
|  | 474 | III | ZM | ∅ | ∅ |
|  | 475 | III | FMc | -6.9427 | Nodal line |
|  | 476 | III | Canted-FM | -6.9428 | Weyl points |
|  |  |  | FMa | -6.9428 |  |
|  |  |  | FMb | -6.9428 |  |
|  | 477 | III | ZM | ∅ | ∅ |

To understand the DFT results and the symmetry protection, we construct an effective model of low-energy bands near $E_F$ around $\Gamma$. In the absence of SOC, the two low-energy bands belong to $\Gamma_1^+$ and $\Gamma_3^-$ irreps (labeled by the point group $D_{2h}$ of $\Gamma$). Thus, we construct the $4 \times 4$ Hamiltonian under the basis of $\{\uparrow, \downarrow\} \otimes \{\Gamma_1^+, \Gamma_3^-\}$, according to the theory of invariants (see details in the Supplementary Information). It reads

$$H(\boldsymbol{k}) = \sigma_0 \otimes \begin{pmatrix} m_1 + m_1^i k_i^2 & 0 \\ 0 & m_2 + m_2^i k_i^2 \end{pmatrix}$$
$$+ t_x k_x \sigma_0 \otimes \tau_y + t_y k_y \sigma_z \otimes \tau_x + t_z k_z \sigma_y \otimes \tau_z$$

of which the parameters are listed in SI. 3, and where $\sigma_i(\tau_i)$ are Pauli matrices in the spin (orbital) space. Zeeman's coupling Hamiltonian along an arbitrary direction is given by $H_B(\theta, \phi)$, which reads

$$H_B(\theta, \phi) = \sigma^{[\theta,\phi]} \otimes \begin{pmatrix} g_1 & 0 \\ 0 & g_2 \end{pmatrix}, \quad \sigma^{[\theta,\phi]} \equiv \begin{pmatrix} \cos(\theta) & e^{-i\phi}\sin(\theta) \\ e^{i\phi}\sin(\theta) & -\cos(\theta) \end{pmatrix}$$

where $g_1 = -0.135$ eV and $g_2 = 0.105$ eV, the polar angle $\theta$ and azimuthal angle $\phi$ indicate the direction of the magnetism (e.g., $\theta = 0$, $\sigma^{[\theta,\phi]} = \sigma_z$ for the z-direction). With an external magnetic field, the FM states along different directions can be achieved. The $H(\boldsymbol{k}) + H_B(\theta, \phi)$ Hamiltonian yields the same results as the DFT simulations. The nodal line in FMc is protected by $m_z$ symmetry, while the Weyl points is protected by two-fold rotation in the FMa/FMb states.

The two bands around Fermi level $E_F$ at $\Gamma$ in the absence of spin-orbit coupling (SOC) $\Gamma_1^+$ and $\Gamma_3^-$ (labeled according to the point group $D_{2h}$). The basis set is then chosen to be $\{\uparrow, \downarrow\} \otimes \{\Gamma_1^+, \Gamma_3^-\}$. The $\boldsymbol{k} \cdot \boldsymbol{p}$ Hamiltonian at $\Gamma$ is constrained by point group $D_{2h}$ (the generators are $C_{2z}$, $C_{2y}$, and $P$) and time-reversal symmetry T. In such basis setup, the matrix representations of these symmetry operations are

$$\mathcal{D}(C_{2z}) = -i\sigma_z \otimes \sigma_0$$
$$\mathcal{D}(C_{2y}) = -i\sigma_y \otimes \sigma_z$$
$$\mathcal{D}(P) = \sigma_0 \otimes \sigma_z$$
$$\mathcal{D}(T) = -i\sigma_y \otimes \sigma_0 \widehat{K}$$

Where $\widehat{K}$ is the complex conjugation operator. According to the theory of invariants $[\mathcal{D}(g)H(\boldsymbol{k})\mathcal{D}(g)^\dagger = H(g\boldsymbol{k})]$, the $4 \times 4$ effective Hamiltonian is constructed as below,

$$H(\boldsymbol{k}) = \sigma_0 \otimes \begin{pmatrix} m_1 + m_1^i k_i^2 & \\ & m_2 + m_2^i k_i^2 \end{pmatrix}$$
$$+ t_x k_x \sigma_0 \otimes \sigma_y + t_y k_y \sigma_z \otimes \sigma_x + t_z k_z \sigma_y \otimes \sigma_z$$

of which the parameters are listed in SI 4.

The Zeeman's coupling is obtained as,

$$H'(\boldsymbol{B}) = \mu_B B^i \sigma_i \otimes \begin{pmatrix} g_1 & \\ & g_2 \end{pmatrix}$$

where $i = x, y, z$, $\mu_B = \frac{e\hbar}{2m_e}$ is the Bohr magneton, $g_1 = -0.135\ T^{-1}$ and $g_2 = 0.105\ T^{-1}$ are the fitted $g$ factors. Under in-plane ($xy$ plane) magnetic field, or in the FM state along $a/b$, $H$ is a magnetic Weyl semimetal hosting a single Weyl point on the $\Gamma - X / \Gamma - Y$ high symmetry line. While under the external magnetic field along $c$, or in the FM state along $c$, $H$ is a nodal-line semimetal, hosting a nodal ring in the $k_x k_y$ plane.

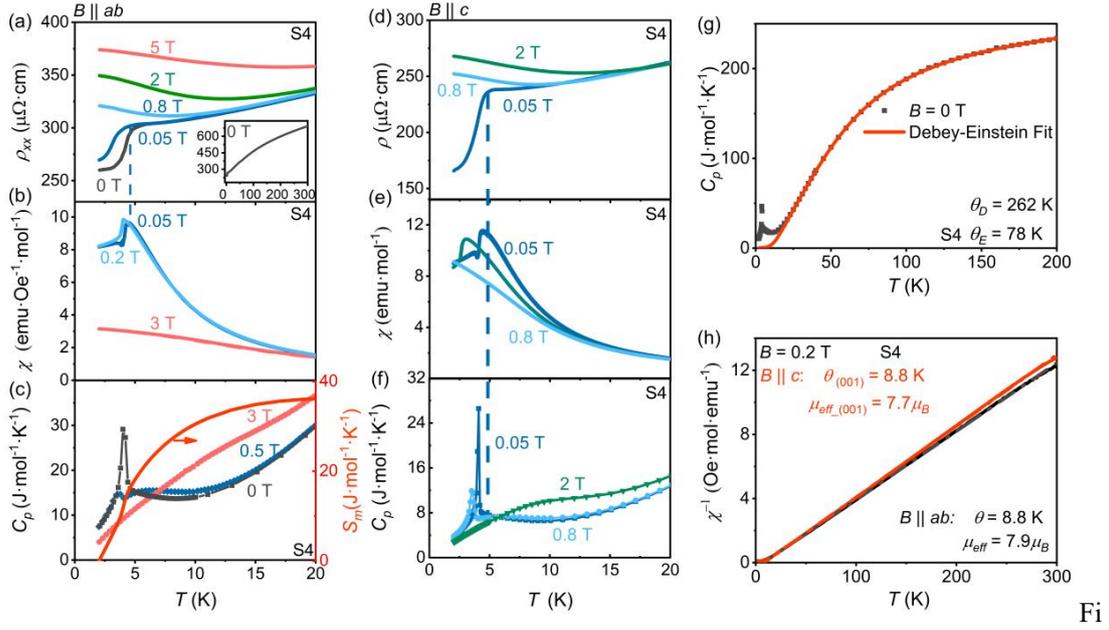

Fig.2 (a) - (c) and (d)-(f) depict the temperature dependence of resistivity, magnetic susceptibility, and specific heat of Eu$_3$In$_2$As$_4$ for $B\|ab$ and $B\|c$, while the red line in (c) represents the magnetic entropy at $B = 0$ T; the insets in (a) is the resistivity in range of 2 - 300 K. (g) is the fit of specific heat by Debye-Einstein; (h) is the result of Curie-Weiss fit.

*Magnetism and Weyl semi-metallic state.* The resistivity of Eu$_3$In$_2$As$_4$ decreases with cooling [see the inset of Figure 2(a)], revealing a metallic nature at 0 T. Below 5 K, resistivity shows a sharp drop, which is possibly related to the AFM order. The AFM order is further proved by peak in susceptibility curves $\chi(T)$ and λ-shape peak in specific heat. The $\chi^{-1}(T)$ data, between 10 – 300 K, can be fitted with the Curie-Weiss law $\chi = \frac{C}{T-\theta}$, where the $C$ and $\theta$ are Curie constant and Weiss temperature, respectively. The fitting results in $\theta = 8.8$ K for both direction and $C = 2.35$ and 2.24 emu·K·Oe$^{-1}$·mol$^{-1}$ for $B \| ab$ and $B \| c$, respectively. The corresponding effective moment $\mu_{eff}$ is 7.9 $\mu_B$ and 7.7 $\mu_B$, close to theoretical value of Eu$^{2+}$ (S=7/2, $\mu_{eff} = 7.9\ \mu_B$) [Fig.2 (g)]. In Figure 2(a), the resistivity of Eu$_3$In$_2$As$_4$ decreases as the temperature decreases below 5 K, indicating a metallic behavior. It contradicts the estimated small energy gap (4.4 meV) in the DFT caluculations, which may be due to the shift of Fermi level in the samples.

By subtracting the phonon contributions [Fig.2 (g)], we derive the entropy 36.2

J·mol⁻¹·K⁻¹ contributed by magnetic $Eu^{2+}$ ions [Fig.2 (c)]. It is smaller than the theoretical value of 3RIn8(R is universal gas constant), which may be due to the overestimation of the phonon component from the Debye-Einstein fitting.

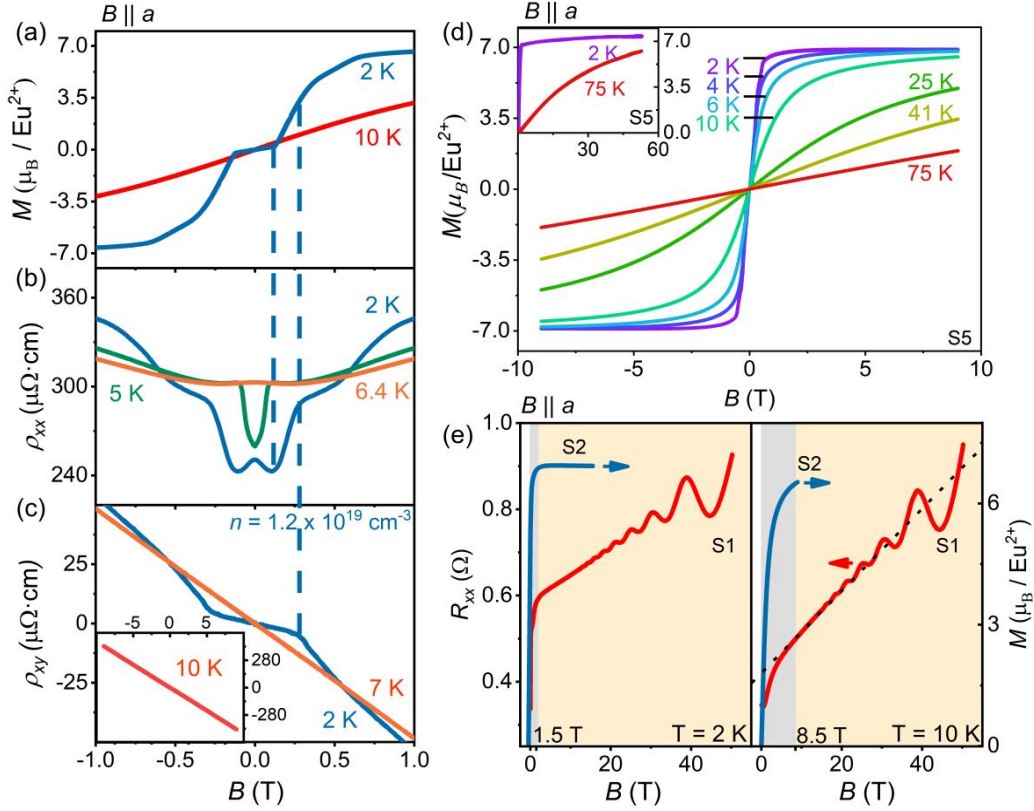

Fig.3 (a) depict magnetization curve for $B \parallel a$; (b) and (c) are magnetoresistivity and Hall resistivity for $B \parallel a$, inset illustrates a linear behavior range from -9 T – 9 T in Hall resistivity; (d) is the magnetization $M(T)$ under different temperatures for $B \parallel a$; (e) is comparations between magnetoresistance and magnetization for T = 2 K and T = 10 K, respectively. The data demonstrates that $Eu_3In_2As_4$ is in Weyl topological state at field-induced FM-state.

As shown in Fig.3 (a), the field-dependent magnetization curves, $M(B)$, show linear behavior around zero fields, and no hysteresis is observed. As the magnetic field increases, $M(B)$ curve measured at 2 K displays slope changes around 0.2 T and 0.5 T. As in Fig.3 (d) and (e), the magnetization saturates around 1.5 T at $T = 2$ K for $B \parallel a$ ($7.9\mu_B/Eu^{2+}$), where $Eu_3In_2As_4$ is in a field-induced FM state. These field-induced transitions are corresponding to the changes shown in magnetoresistivity for $B \parallel a$ (Fig.3 (b) and (c)). Similar transitions are observed for $B \parallel c$ as in SI. 5.

Fig.3 (c) shows linear Hall resistivity with anomalous Hall behavior around 0 T, which is depressed gradually with the temperature increasing. Hall resistivity can be described by: $\rho_H = R_H B + R_S \mu_0 M$, where the $R_H, B, R_S, \mu_0$ and $M$ represent normal Hall coefficient, intensity of magnetic field, abnormal Hall coefficient, magnetic flux constant, and magnetic moment. The second term represents the effect of magnetic atoms. The anomalous Hall of $Eu_3In_2As_4$ is notably weaker than the typical Weyl semimetal at low magnetic fields. Given that Fermi arcs contribute to the anomalous Hall effect, a tiny anomalous Hall effect could indicate a close proximity between the Weyl points. We use $n = \frac{1}{R_H}$ to simplify the process of calculation, neglecting the tiny effect caused by magnetization. $R_H$ is independent of the magnetic field for one band model[34, 35]. The model indicates a linear Hall conductivity behavior and the best fit yields the density of electrons carriers $n = -1.2 \times 10^{19}$ cm$^{-3}$.

The Weyl semimetallic state in $Eu_3In_2As_4$ is found to be intimately connected to the ferromagnetic state shown in Fig 3.(e). As is well-known, the unsaturated linear behavior is a characteristic feature of the Weyl semimetal. We use this characteristic as an indicator to identify the magnetic field where $Eu_3In_2As_4$ is in the Weyl semimetallic state. In fact, we observed that the magnetization of $Eu_3In_2As_4$ saturates at the same intensity of the magnetic field where $Eu_3In_2As_4$ exhibits the Weyl semimetallic state in Fig 3. (e). Fig.3(d) and Fig.4(c) (or SI.8) exhibit a correlation between saturated magnetization and topological state, demonstrated by a discernible shift in oscillation upon unsaturation of magnetization at or above 25 K.

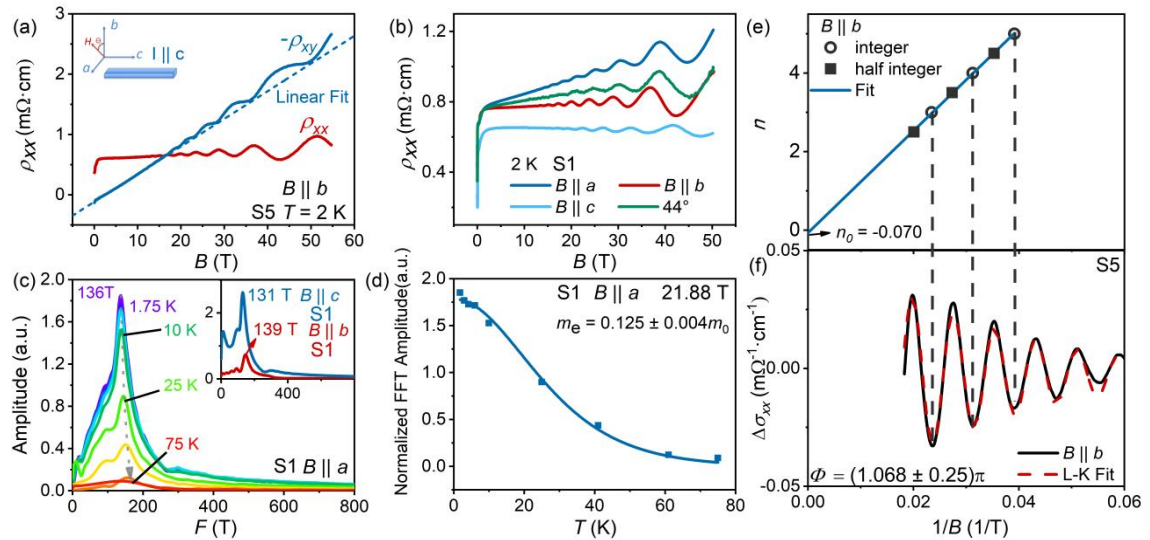

Fig.4 (a) and (d) show the original data of the dependence of the field. (b) and inset in (b) are the results of FFT analysis for different direction and (e) is the result of $R_T$; (c) and (g) Landau Level (LL) index fan diagram to get the intercept $n_0$ on n axis and L-K fitting for oscillation respectively.

*Quantum oscillation.* Furtherly, we use a maximum magnetic field of up to 54.7 T to measure the *MR* of Eu$_3$In$_2$As$_4$. As shown in Fig.4 (a) and (b), the MR curves above 14 T exhibit a strong oscillation with only one frequency. As no-go theorem, Weyl points always appear in pair, which seems to be conflict with the single frequency. However, projected area of the two Weyl points along specific directions are equal, leading to closely spaced oscillation frequencies. Linear unsaturated magnetoresistivity is also observed for $B \parallel a$ and $B \parallel b$ as an indicative characteristic of WSM.

Fig. 4 (c) shows a deviation in the peak position at 75 K compared to 1.8 K, indicating changes in the Fermi surface as temperature rises. A temperature increase from 10 K to 25 K may significantly disrupt the formation of a field-induced topological state, as evidenced by a sharp change in FFT. A shoulder adjacent to the peak was observed. However, it was absent during a retesting measurement [in SI.5]. Given not any obvious additional frequency is observed beside the main peak [in SI.6], we treat it as single frequency along *a/b* axis. For *c* axis, we cannot exclude the possibility of the shoulder being an intrinsic frequency component. The result shows a peak *F* at 137 T, 147 T, 131 T for $B \parallel a$, $B \parallel b$ and $B \parallel c$, corresponding to external cross-section area on Fermi surface $A_F$ = 1.29 nm$^{-2}$, 1.39 nm$^{-2}$ and 1.24 nm$^{-2}$ respectively by using the Onsager relation $F = (\Phi_0/(2\pi^2))A_F$, where $\Phi_0 = 2.07 \times 10^{-15}$ T·m$^2$ is unit of quantum magnetic flux. We deduce the mobility of Eu$_3$In$_2$As$_4$ by assuming the shape of a Fermi pocket as an ellipsoid. We obtained the Fermi vectors $k_{F\_a} = 0.64\ nm^{-1}$, $k_{F\_b} = 0.66\ nm^{-1}$, and $k_{F\_c_\alpha} = 0.62\ nm^{-1}$ , which are pretty close to each other. Using the three-dimensional electron gas model, $k_F^3 = 3\pi^2 n$, where the *n* represents the electronic density of the material, we get a $n_a \approx 0.89 \times 10^{19}\ cm^{-3}$, $n_b \approx 0.97 \times 10^{19}\ cm^{-3}$ and $n_{c\_\alpha} \approx 0.80 \times 10^{19}\ cm^{-3}$ which is close to the Hall result. This finding suggests that almost all carriers in the single band contributing to the Hall conductivity participate in cyclotron motion, providing confirmation of the presence of only one pair of Weyl points around the Fermi level.

The mobility of electrons in Eu$_3$In$_2$As$_4$ is evaluated by $ne\mu = \sigma_0$, where $\mu$ represents mobility and $\sigma_0$ represents conductivity at around $B = 0$ T. From the formula, we obtained $\mu_a = 2300\ cm^2 \cdot V^{-1} \cdot s^{-1}$, $\mu_b = 1600\ cm^2 \cdot V^{-1} \cdot s^{-1}$ and $\mu_{c\_\alpha} = 3900\ cm^2 \cdot V^{-1} \cdot s^{-1}$ at 2 K. Considering that Eu$_3$In$_2$As$_4$ is a nodal line semimetal in the case of B∥c, the absence of negative magnetoresistivity is reasonable ($I\|c$)[36].

As shown in Fig.4(d), the cyclotron mass m* on Fermi surface is obtained generally from the fit of the temperature dependence of FFT amplitude, using the temperature damping factor of the Lifshitz–Kosevich (L-K) equation[37], $R_T = \frac{\alpha T m^*}{m_0 B \sinh(\frac{\alpha T m^*}{m_0 B})}$, in which $\alpha = \frac{2\pi^2 k_B m_0}{\hbar e} \approx 14.69\ T \cdot K^{-1}$, $T$, $B$, $m_0$, $m^*$ and $\hbar$ represent the measurement temperature, the average field, effective mass of electron, rest mass of the electron and Plank constant respectively. The average $B$ is determined from $((B_{start}^{-1} + B_{end}^{-1})/2)^{-1} = 21.975$ T with $B_{start} = 14$ T and $B_{end} = 50$ T. In the case of $B \parallel a$, a relatively small cyclotron mass $m^* \approx 0.125 \pm 0.004\ m_0$ is obtained.

From Fig. 4(a), $\rho_{xx}$ is close to $\rho_{xy}$. Therefore, the electronic conductivity data, obtained by $\sigma_{xx} = \frac{\rho_{xx}}{\rho_{xx}^2 + \rho_{xy}^2}$ for $B \parallel b$ is used to analyze oscillation[38]. To fit the oscillatory component, we employ the standard L-K expression[37]:

$$\frac{\Delta\sigma_{xx}}{\sigma_{bg}} = \sqrt{\frac{B}{F}} \cdot e^{-\alpha T_D m_e^*/B} \cdot \frac{\alpha T m^*}{m_0 B \sinh\left(\frac{\alpha T m^*}{m_0 B}\right)} \cdot \cos\left(\frac{2\pi F}{B} + \varphi + 2\pi\delta\right)$$

where the $\Delta\sigma_{xx}$, $\sigma_{bg}$, $T_D$ and $\varphi$ represent the oscillatory component, the background, Dingle temperature, and Berry phase. We obtain a reasonable fit to the observed oscillations (bold curve) using just one frequency as shown in Fig. 4(f). The optimal fit yields $\varphi + 2\pi\delta = 1.068\pi$, and $T_D = 55$ K. For a 3D system limit, the factor $|\delta| = 1/8$, representing a nontrivial $\pi$ Berry phase. The Dingle temperature reflects the effect of electron scattering on the reduction of the oscillation amplitude. This effect is always noted as the formation of electronic relaxation time described as $\tau = \frac{\hbar}{2\pi k_B T_D} = 0.02\ ps$, indicating a rather small effect of relaxation time on oscillation reduction in Eu$_3$In$_2$As$_4$.

For Dirac or Weyl semimetal, the non-trivial Berry phase could also be accessed from either Landau level (LL) index fan diagram. As in Fig. 4(e), The LL integer index n is assigned to the oscillatory minimum of $\sigma_{xx}$, while the half-integer index $n+1/2$ is assigned to the oscillatory maximum of $\sigma_{xx}$. The intercept $n_0$ on the $n$ axis of the LL fan diagram is -0.070, which gives $\varphi_{ex} = 2\pi \left(n_0 + \frac{1}{2}\right) = 0.86\pi$ in Fig. 4(e). Taking the factor δ into account, the result gives strong evidence for non-trivial π Berry phase $\Phi_B$.

In summary, we have synthesized single crystal $Eu_3In_2As_4$. The DFT results have revealed a single pair of Weyl points in FMa/FMb/canted-FM states. Magnetic measurements reveal that $Eu_3In_2As_4$ shows a field-induced FM state above $H = 1.5$ T at 2 K. Magnetoresistivity data indicate a single frequency sdH oscillation for $Eu_3In_2As_4$ for specific *a/b* axis. The linear field dependence of the Hall conductivity is consistent with a one-band model. The close value of the carrier density derived from Hall resistance and oscillation suggests that all carriers contributing to the Hall conductivity participate in cyclotron motion. Furthermore, the analysis of the oscillation phase indicates a π Berry phase. All these results indicating that $Eu_3In_2As_4$ is a potential MWS with a single pair of Weyl points under magnetic fields.

*Note: When completed this article, we became aware of related work[39].*

## Acknowledgments

This work is supported by the National Natural Science Foundation of China (12104492, U2032204, 92065203, and 12174430), National Key Research and Development Program of China (2018YFA0702100, and 2021YFA1400401), the Strategic Priority Research Program B of the Chinese Academy of Sciences (XDB33000000), the Beijing Nova Program (Z211100002121144), the China Postdoctoral Science Foundation (2021TQ0356), the Center for Materials Genome, and the Synergetic Extreme Condition User Facility (SECUF).

**Author contributions:** Ke Jia prepared the single crystals under the supervision of Youguo Shi. Ke Jia performed the single crystal x-ray diffraction. The magnetic and transport measurements in the PPMS-9T were carried out by Ke Jia, Yupeng Li, Dayu Yan, Cuixiang Wang, Hai L. Feng, and Jie Shen. The high-field

# Discovery of a Magnetic Topological Semimetal Eu$_3$In$_2$As$_4$ with a Single Pair of Weyl Points

Ke Jia[1,2,3#], Jingyu Yao[1,3#], Xiaobo He[4#], Yupeng Li[1,3#], Junze Deng[1,3], Ming Yang[4], Junfeng Wang[4], Zengwei Zhu[4], Cuixiang Wang[1], Dayu Yan[1], Hai L. Feng[1,5], Jie Shen[1,2,3,5*], Yongkang Luo[4*], Zhijun Wang[1,3*], Youguo Shi[1,2,3,5*]

Supplementary information contains four tables I-VI in SI.1-SI.4 and six Figures SI.5-SI.10, detail of calculation is in SI.11 and details of experiment is in SI.12

SI.1-2 The result of Single-Crystal X-ran Diffraction of Eu$_3$In$_2$As$_4$

SI.3 The fitted parameters for the $\bm{k} \cdot \bm{p}$ Hamiltonian;

SI.4 Possible magnetic structure of Eu$_3$In$_2$As$_4$;

SI.5 The magnetoresistivity and the first derivation of magnetization;

SI.6 Magnetoresistance of multiple samples for $H \parallel ab$ and $H \parallel c$;

SI.7 Retest sdH oscillation on sample 5 with magnetic field up to 55 T;

SI.8 The oscillatory component under series temperature;

SI.9 Hall with magnetic field up to 55 T;

SI.11 Details of experiments;

| | |
|---|---|
| Chemical formula | $Eu_3In_2As_4$ |
| Formula weight | 985.20 g/mol |
| radiation | Mo $K\alpha$, 0.71073 Å |
| Temperature | 273.(2) K |
| Crystal system | orthorhombic |
| Space group | *P nnm* |
| Unit cell dimensions | $a$ = 6.8300(4) Å<br>$b$ = 16.5067(9) Å<br>$c$ = 4.4102(2) Å |
| volume | 497.21(5) Å$^3$ |
| Z | 2 |
| Density(calculated) | 6.581 g/cm$^3$ |
| Absorption coefficient | 38.753 mm$^{-1}$ |
| F(000) | 838 |
| Theta range for data collection | 2.4678° to 36.3777° |
| Index ranges | $-11 \leq h \leq 11$<br>$-27 \leq k \leq 27$<br>$-7 \leq l \leq 7$ |
| Independent reflections | 2123[R(int)=0.0816] |
| Structure solution program | SHELXT 2014/5 |
| Refinement program | SHELXT-2018/3 |
| Function minimized | $\Sigma W(F_o^2-F_c^2)^2$ |
| Goodness of fin on F$^2$ | 1.064 |
| R1 (I > 2σ$_1$) | 0.0697 |
| ωR2(I > 2σ$_1$) | 0.1754 |
| R1 (all data) | 0.0700 |
| ωR2(all data) | 0.1758 |
| Weighting scheme | $W$=1/[σ$^2$(F$_o^2$)+(0.0146P)$^2$+0.1060P]<br>Where P = ( F$_o^2$+2F$_c^2$) |

SI.1. Table I. Crystallographic data of Eu$_3$In$_2$As$_4$, the result has been reported in Ref.32.

|      | x/a         | y/b         | z/c  | U(eq)        |
|------|-------------|-------------|------|--------------|
| Eu1  | 1.00        | 0.50        | 1.00 | 0.00720(18)  |
| Eu2  | 0.70821(8)  | 0.69827(3)  | 0.00 | 0.00966(18)  |
| In   | 0.36038(9)  | 0.58630(4)  | 0.50 | 0.00719(19)  |
| As1  | 0.74604(14) | 0.56970(6)  | 0.50 | 0.00660(2)   |
| As2  | 0.23329(15) | 0.66851(6)  | 0.00 | 0.0074(2)    |

SI.2. Table II. Atomic coordinates and equivalent isotropic thermal parameters of the Eu$_3$In$_2$As$_4$

| 0th order | eV | 1st order | eV·Å | 2nd order | eV·Å² |
|---|---|---|---|---|---|
| $m_1$ | 0.096315 | $t_x$ | 1.8 | $m_1^x$ | 41 |
| $m_2$ | -0.09644 | $t_y$ | 3.6 | $m_1^y$ | 21 |
| — | — | $t_z$ | 4.3 | $m_1^z$ | 8 |
| — | — | — | — | $m_2^x$ | -0.8 |
| — | — | — | — | $m_2^y$ | -1 |
| — | — | — | — | $m_2^z$ | -10 |

SI.3 Table IV. The fitted parameters for the $\mathbf{k}\cdot\mathbf{p}$ Hamiltonian.

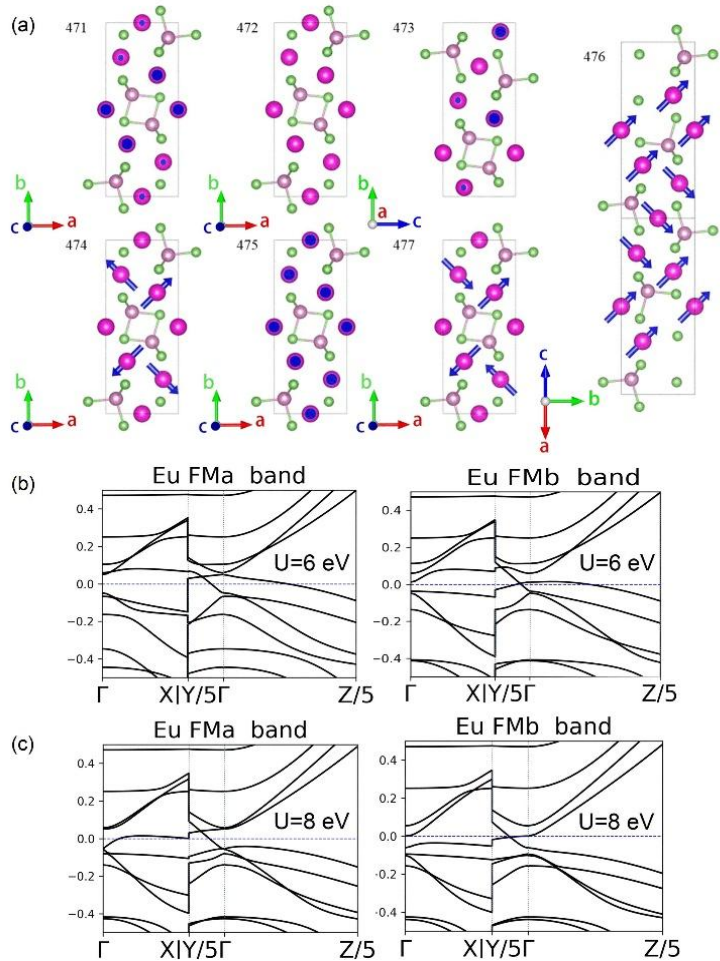

SI.4 (a)The initial magnetic configurations of $Eu_3In_2As_4$ in MSG #471-477. Both reasonable and unreasonable structures are shown. Red circles represent Eu atoms, pink circles represent In atoms, and green curcles represent As atoms. Blue arrows represent the nonzero magnetic moments. The moment of some Eu atoms is confined to zero by MSG, which are unreasonable; (b) and (c) illustrate the band structure of $Eu_3In_2As_4$ with U = 6 eV and 8 eV, respectively.

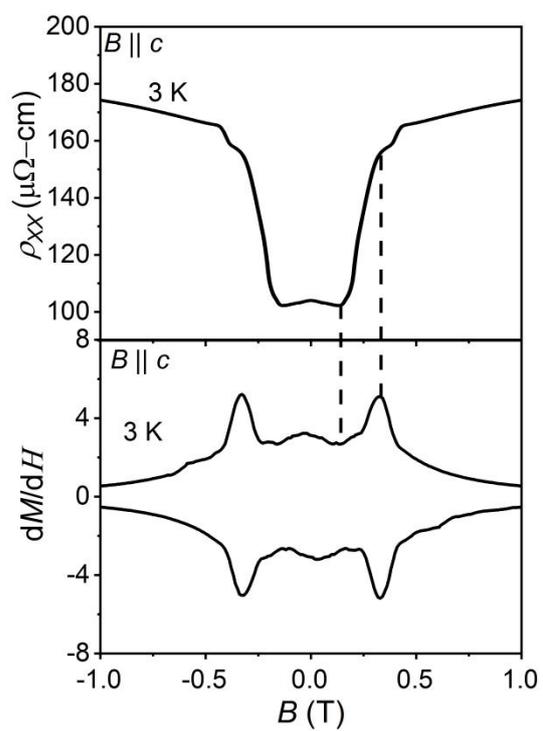

SI.5 The magnetoresistivity and the first derivation of magnetization. The transition induced by magnetic field is similar to *B* || *ab*.

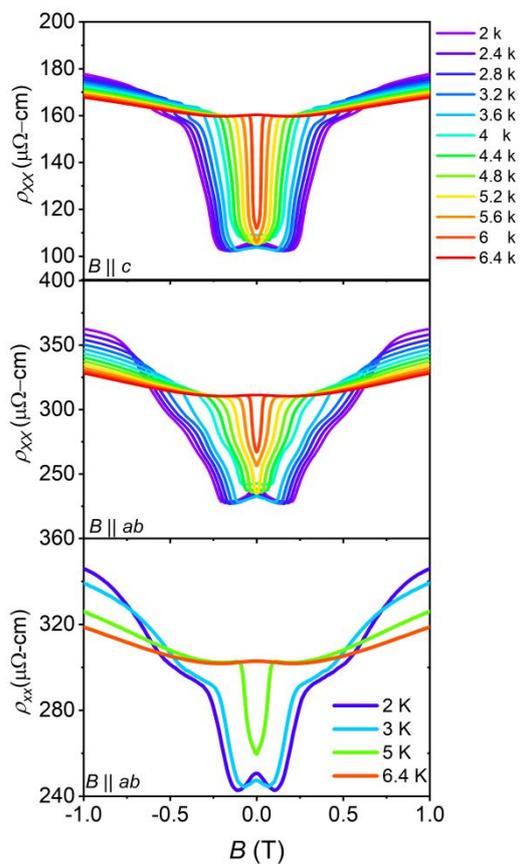

SI. 6 We have measured magnetoresistance of multiple samples of Eu$_3$In$_2$As$_4$. Though some details differ from sample to sample, the main characteristics are absolutely similar to each other.

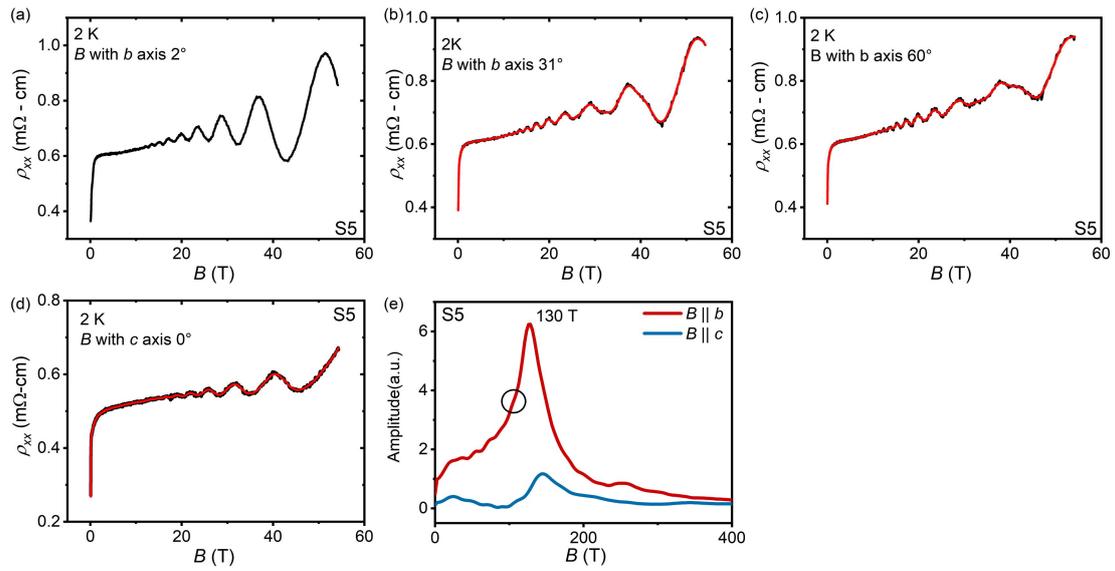

SI.7 We retested the magnetoresistivity with changing the angle misaligned [010] from 2° to 90° with instantaneous magnetic field up to 54 T. However, the signals are not ideal except *b* direction. The FFT analysis proved the fake peak mentioned above in SI.6 (e), further comfirming a single frequency of sdH oscillation.

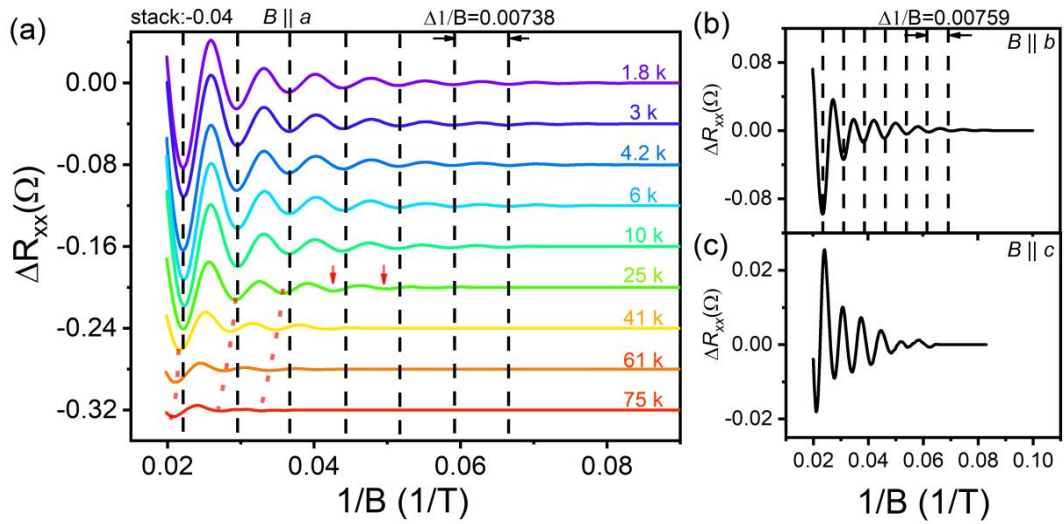

SI.8 The oscillatory component under series temperature from 1.8 K to 75 K for $B \parallel a$ and the oscillatory component under 1.75 K for $B \parallel b$ and $B \parallel c$; (a) demonstrates an obvious shift as temperature increases by dotted lines and arrows.

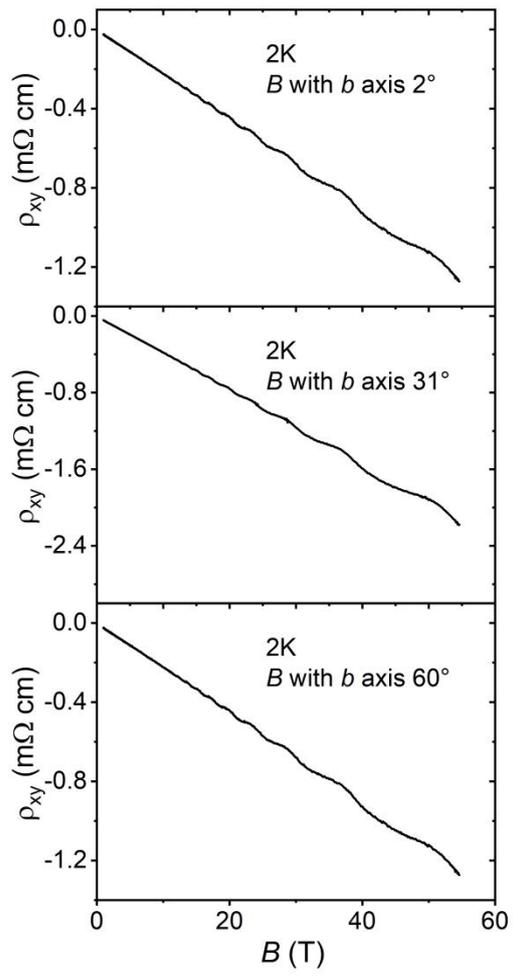

SI.9 Hall measurement along different directions at *ab*-plane.

SI.10 Details of method

The Eu$_3$In$_2$As$_4$ was prepared by using In and Sn as mixed flux. Elements Eu (> 99.9%, Alfa Aesar), Sn(4N, Alfa Aesar), In (4N, Alfa Aesar), and As (4N, Alfa Aesar), with a mole ratio of 3:15:15:4.2, were loaded into an alumina crucible. Then the crucible was placed into a silica tube. All these processes were carried out in an argon-filled glove box to protect the elements. The silica tube was transferred outside of the glovebox and sealed under a dynamic vacuum using a C$_2$H$_2$/O$_2$ torch, at 1000 °C for 10 h, then cooled slowly (2 °C/h) to 800 °C. After dwelt for 5 hours at 800°C, the sealed tube was centrifuged immediately to remove the excess molten Indium. Needle-shaped single crystals of Eu$_3$In$_2$As$_4$ up to several millimeters in length were obtained.

The composition of the single crystals was checked using energy dispersive x-ray spectroscopy (EDX) operated in a Hitachi S-4800 scanning electron microscopic at an accelerating voltage of 15 kV and an accumulation time of 90 seconds.

The temperature dependence of magnetic susceptibility, $\chi$(T), was measured in a SQUID VSM-7T (Quantum Design Inc.) between 2 K and 300 K with varied applied magnetic fields parallel (or vertical to c axis) of Eu$_3$In$_2$As$_4$ crystals under field-cooled (FC) and zero-field-cooled (ZFC) conditions. Isothermal magnetization curves were measured at 2 K, 10 K, and 300 K with applied magnetic fields between -9 to 9 T. The temperature dependence of specific heat $C_p$(T) and resistivity R$_{xx}$(T) was measured in a physical property measurement system (PPMS-9 T, Quantum Design Inc.) from 2 K to 300 K with varied magnetic fields parallel and vertical to b axis, respectively. Both the magnetic dependence of resistivity $\rho_{xx}$(H) and the Hall $\rho_{xy}$(H) are in PPMS mentioned above from -9 T to 9 T under the varied temperatures. The standard four-probe method was employed for the resistivity measurements. High-field MR and Hall effect data up to 54.7 T were obtained by pulse magnetic field at Wuhan National High Magnetic Field Center.

Noncollinear DFT+U+SOC were calculated by the frozen-core projector augmented plane wave (PAW) encoded in the Vienna ab Initio Simulation Package (VASP) and the PBE exchange-correlation functional. The electron correlation associated with the 4f states of Eu was taken into consideration by DFT+U calculations with an effective on-site repulsion U = 7 eV. All our DFT calculations used the plane wave cutoff energy of 500eV, the set of (8×4×12) k-points, and the threshold of 1E-6 eV for self-consistent-field energy convergence.